\documentclass[12pt,letter,final]{iopart}
\usepackage{iopams}  
\usepackage{graphicx}
\usepackage{epstopdf}
\usepackage[breaklinks=true,colorlinks=true,linkcolor=blue,urlcolor=blue,citecolor=blue]{hyperref}
\usepackage[numbers,sort&compress]{natbib}

\begin{document}
\title[Sympathetic cooling of molecular ion motion to the ground state]{Sympathetic cooling of molecular ion motion to the ground state}

	\author{R Rugango$^{1}$, J E Goeders$^{1}$\footnote{Current address: Honeywell ACS Labs, 1985 Douglas Drive North, Golden Valley, MN 55422}, T H Dixon$^{2}$, J M Gray$^{1}$\footnote{Current address: Department of Physics, University of Colorado, Boulder, CO 80309}, N Khanyile$^1$, G Shu$^1$, R J Clark$^3$, and K R Brown$^1$} 
	 \address{$^{1}$Schools of Chemistry and Biochemistry, Computational Science and Engineering, and Physics, Georgia Institute of Technology, Atlanta, GA 30332}
	\address{$^2$Department of Chemistry, Louisiana Tech University, Ruston, LA 71272} 
	\address{$^3$Department of Physics, The Citadel, Charleston, SC 29409}
	\ead{kenbrown@gatech.edu}
    
\begin{abstract}
We demonstrate sympathetic sideband cooling of a $^{40}$CaH$^{+}$ molecular ion co-trapped with a  $^{40}$Ca$^{+}$ atomic ion in a linear Paul trap. Both axial modes of the two-ion chain are simultaneously cooled to near the ground state of motion. The center of mass mode is cooled to an average quanta of harmonic motion $\overline{n}_{\mathrm{COM}} = 0.13 \pm 0.03$, corresponding to a temperature of $12.47 \pm 0.03 ~\mu$K. The breathing mode is cooled to $\overline{n}_{\mathrm{BM}} = 0.05 \pm 0.02$, corresponding to a temperature of $15.36 \pm 0.01~\mu$K.
\end{abstract}

\vspace{2pc}


\section{Introduction}

Trapped, cold molecular ions are a promising system for measuring fundamental physical constants from the determination of the electron electric dipole moment \cite{LohScience2013} to the time variation of the electron-to-proton mass ratio \cite{KajitaJPhysB2011,BresselPRL2012}. Precision spectroscopy requires molecular ions with internal degrees of freedom prepared in specific states and cold external degrees of freedom to reduce unwanted Doppler shifts. For specific molecular ions, this may be achievable by direct laser cooling \cite{NguyenPRA2011,NguyenNJP2011}, as has been the case for neutral molecules \cite{ShumanNature2010,HummonPRL2013,ZhelyazkovaPRA2014}, but for most molecular ions a combination of cooling methods is required.

The internal degrees of freedom have been controlled using state-selective photoionization \cite{TongPRL2010}, optical pumping \cite{StaanumNatPhys2010,SchneiderNatPhys2010,LienNatComm2014}, buffer gas cooling \cite{HansenNature2014}, and sympathetic cooling with laser-cooled neutral atoms \cite{RellegertNature2013}.   The external degrees of freedom have been cooled by buffer gas cooling \cite{BoyarkinJACS2006} and sympathetic cooling of co-trapped atomic ions \cite{MolhavePRA2000,BlythePRL2005,RyjkovPRA2006,WillitschPCCP2008}.  Many sympathetic cooling experiments involve large ion crystals where the translational temperature of the ions is determined by a competition between laser-cooling and the RF driven micromotion.  The RF driven micromotion can be eliminated by aligning an ion chain with the null of the RF trap.

At low temperatures, the motion of ions in an ion chain is best described by normal modes arising from the Coulombic coupling between the ions.  These normal modes can be cooled below the Doppler limit using resolved sideband cooling \cite{MonroePRL1995}. For atomic ions, sympathetic sideband cooling has been used in quantum information experiments \cite{BarrettPRA2003,JostNature2009} and to build precise ion clocks based on quantum logic spectroscopy \cite{wineland,Rosenband2008,ChouPRL2010}. Quantum logic spectroscopy and similar approaches \cite{craig,LinPRA2013,HumePRL2011,WanNatComm2013,roos} require measuring the change in motional energy by observing the control ion fluorescence after the target ion has been excited.

Here we demonstrate the resolved sympathetic sideband cooling of a molecular ion. This is an important step towards applying quantum logic spectroscopy techniques to molecular ions. Previous work in the Drewsen group at Aarhus University has shown the resolved sideband cooling of a single mode of motion to  18.1 $\pm$0.4 $\mu$K \cite{poulsen}.  Here we cool both axial degrees of freedom and effectively reach the quantum limit of molecular ion motion along the trap axis.

\section{Experimental Methods}

\begin{figure*}
\centering
\includegraphics[width=\textwidth]{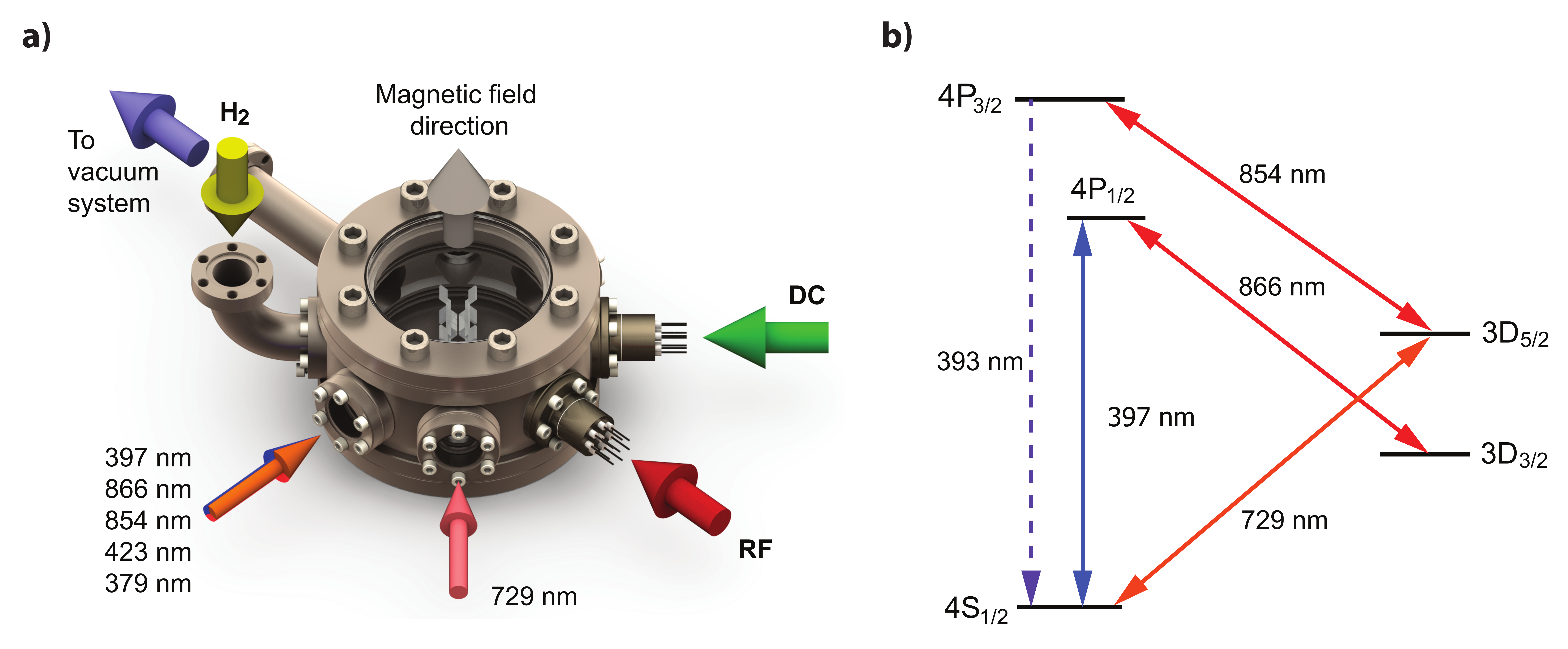}
\caption{(a) The experimental setup including the vacuum chamber and the trap is shown. The magnetic field of 1.4 Gauss is perpendicular to the trap axis and the direction of the lasers. DC connections include compensation voltages, endcap voltages, and oven current. (b)  $^{40}$Ca$^{+}$ level diagram showing the transitions used to  Doppler cool (397 nm, 866 nm) and sideband cool (729 nm, 854 nm) the ions.}
\label{fig:Apparatus}
\end{figure*}

The experiment takes place in a 4.5\verb+"+ spherical octagon vacuum chamber (Kimball Physics MCF450-SphOct) (figure \ref{fig:Apparatus}(a)). The pressure inside the chamber is kept at less than 4$\times$10$^{-9}$ Pa using a 50L/s ion pump (Duniway DSD-050-5125-M) and a Ti sublimation pump (Gamma Vacuum 360819). The chamber houses a five-segment linear Paul trap with $r= 0.5 $~mm previously used for sympathetic heating spectroscopy experiments \cite{craig}. The RF voltage with an amplitude of 122 V oscillates at 14.426 MHz and results in a  secular frequency of 1.419 MHz in the x-direction and 1.475 MHz in the y-direction for $^{40}$Ca$^{+}$  and  a Matthieu \textit{q} of 0.29. All DC voltages are applied through a low pass filter to reduce the RF signal on the DC electrodes. We measure an axial frequency of  568 kHz for $^{40}$Ca$^{+}$  with slighltly unbalanced DC voltages to align the axial RF and DC electric field nulls.

We trap $^{40}$Ca$^{+}$  by evaporating neutral Ca using a stainless steel tube oven and then photoionizing it using 423 nm and 379 nm lasers that are sent into the trap at a 45$^{\circ}$ angle from the trap axis. Both Doppler cooling lasers (397 nm and 866 nm) are coaligned with the photoionization lasers as well as the 854 nm laser used to deshelve the ion during sideband cooling. The 729 nm laser beam used for sideband cooling enters the trap along the axis of the trap. A magnetic field of 1.4 Gauss perpendicular to the axis and the 729 nm polarization splits the Zeeman levels. The relevant  levels of Ca$^+$ are shown in (figure \ref{fig:Apparatus}(b)) and the details of the laser systems can be found in Ref. \cite{goeders}. The ion fluorescence is collected with a lens stack with a numerical aperture of 0.43 and magnification of 10 which sends it to both an  EMCCD camera (Princeton Instruments PhotonMax 512) and a photomultiplier tube (Hamamatsu R.928).

Initially, we trap two $^{40}$Ca$^{+}$ ions and  H$_{2}$ gas is leaked into the vacuum chamber using a  manual leak valve ( Kurt J. Lesker VZLVM967) (figure \ref{fig:Apparatus}a).  The reaction between  $^{40}$Ca$^{+}$ and H$_{2}$ is photoactivated by exciting the  $^{40}$Ca$^{+}$ ion from its 4S$_{1/2}$ ground state to the 4P$_{1/2}$ state \cite{kimura}. A reaction with one of the atomic ions occurs typically 5 min after the addition of H$_{2}$  gas at pressures around 5$\times$10$^{-7}$ Pa. After the reaction , the newly formed dark CaH$^{+}$ molecule is sympathetically cooled through its Coulombic interaction with the remaining  $^{40}$Ca$^{+}$ ion. 

The motional modes of the Coulomb crystal are precisely measured by exciting the 4S$_{1/2}$$\longrightarrow$3D$_{5/2}$ quadrupole transition with the $729$~nm laser and counting the number of electron shelving events \cite{nagerl}. With the 729 nm laser aligned in the axial direction, only the axial motional modes of the crystal can be detected. These modes correspond to a harmonic movement of the two ions in phase, also called the center of mass mode (COM), and out of phase, the breathing mode (BM).

\begin{figure}
\centering
\includegraphics[width=\columnwidth]{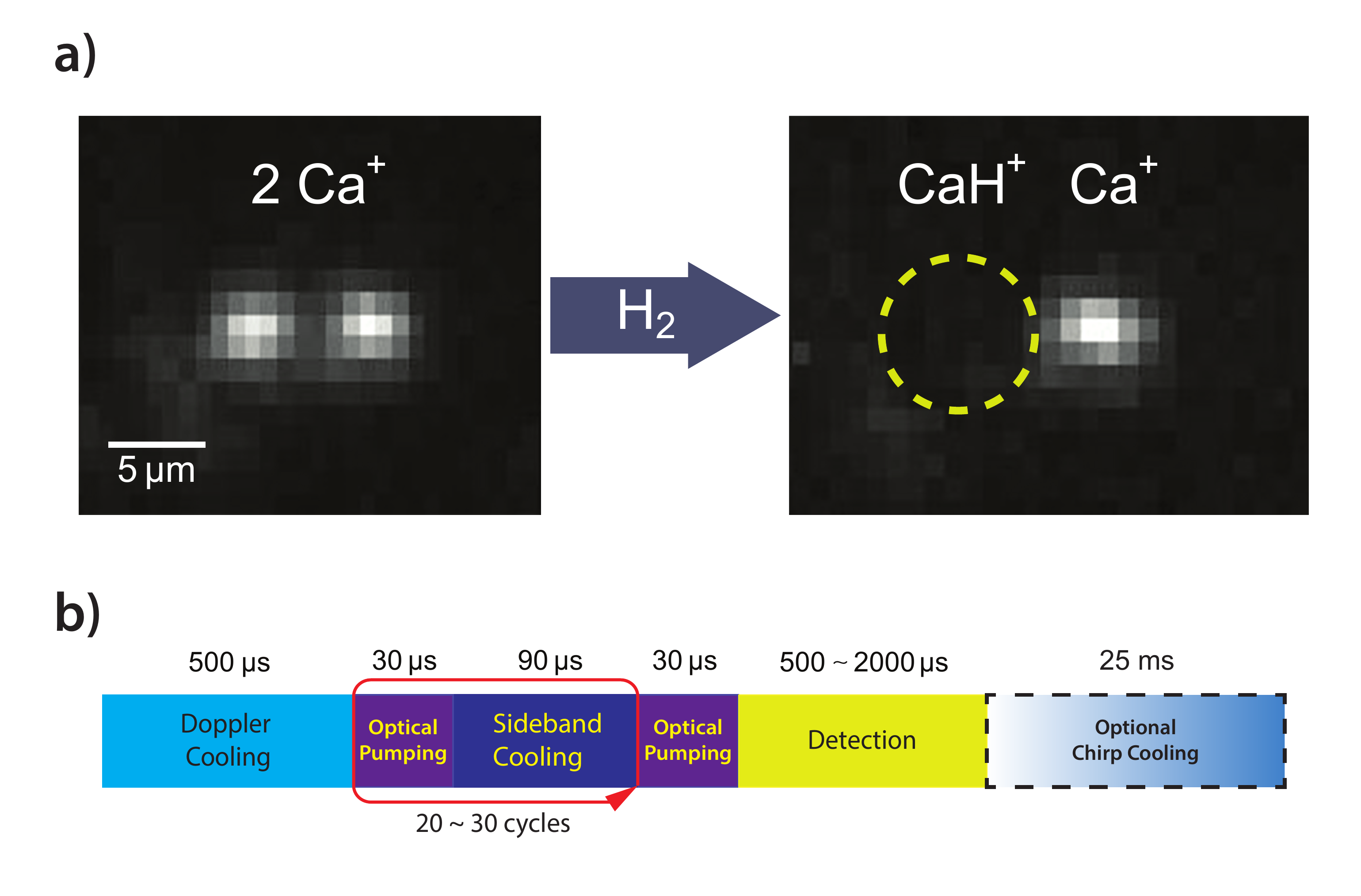}
\caption{(a) One  $^{40}$Ca$^{+}$ reacts with H$_{2}$ to make  CaH$^{+}$ which is not fluorescent. (b) The experiment sequence begins by Doppler cooling the two-ion crystal followed by optical pumping and sideband cooling cycles. After sideband cooling, we perform electron shelving measurements on the motional sidebands and detect the atomic state. If a collision with background gas is detected, then a chirped cooling scheme is applied to recrystallize the ion chain.}
\label{fig:reaction}
\end{figure}

The secular frequencies of the two-ion crystal are related to the axial motional frequency of a  $^{40}$Ca$^{+}$ ion by
\begin{equation}
\nu^{2}_{\pm}=[(1+\mu) \pm \sqrt{1-\mu+\mu^{2}}]\nu_{1}^{2},
\end{equation}
where the $\nu_{-}$ and $\nu_{+}$ correspond to the COM and BM respectively, $\nu_{1}$ is the secular frequency for a single ion, and $\mu$ is the ratio of the mass of the reference atomic ion to the mass of the second ion. This relationship can be used to determine the mass of the molecular ion \cite{DrewsenPRL2004,goeders} and deviations from this relationship can be used to measure stray electric fields \cite{BarrettPRA2003,goeders}. The measured COM frequency is 563$\pm$ 4 kHz while the BM frequency is 976 $\pm$1 kHz. This is in good agreement with the expected values of 564 kHz and 978 kHz and inconsistent with other possible molecular species.

The sideband cooling scheme proceeds by first Doppler cooling the crystal for 500 $\mu$s with the 397 nm laser detuned 10 MHz from resonance and then continuously exciting with the 729 nm laser alternating between the red first order COM and BM sidebands for 6 ms with the 854 nm laser on.  Each cooling cycle is preceded by a  spin polarization phase on the  S$_{1/2}$ (m$_{j}$ = +1/2)$\longrightarrow$D$_{5/2}$ (m$_{j}$ = -3/2) transition to prepare the ion  in the S$_{1/2}$ (m$_{j}$ = -1/2) Zeeman state. Spin polarization is interleaved with the cooling and repeated every 100 $\mu$s. After cooling, we  probe the two-ion crystal with the same technique used to measure motional sidebands (figure \ref{fig:reaction}b). The average motional quanta after cooling is determined through the ratio of the heights of the red and blue sidebands \cite{diedrich}.

The ion crystal occasionally suffers from collisions with residual background gas. Collisions during the sideband cooling procedure will melt the crystal and result in temperatures far above the Doppler cooling limit. Cooling lasers close to resonance may not completely bring the crystal back to temperatures near the Doppler limit within the normal cooling period. Collision events are detected by observing below normal fluorescence during the regular Doppler cooling stage. After detection of a collision,  a frequency chirped Doppler cooling pulse is applied that recrystallizes the ion chain and achieves the desired initial temperature for performing sideband cooling. We discard the electron shelving data recorded coincident with the collision event from our final data set.       

\section{Results and Discussion}
\begin{figure*}
\centering
\includegraphics[width=\textwidth]{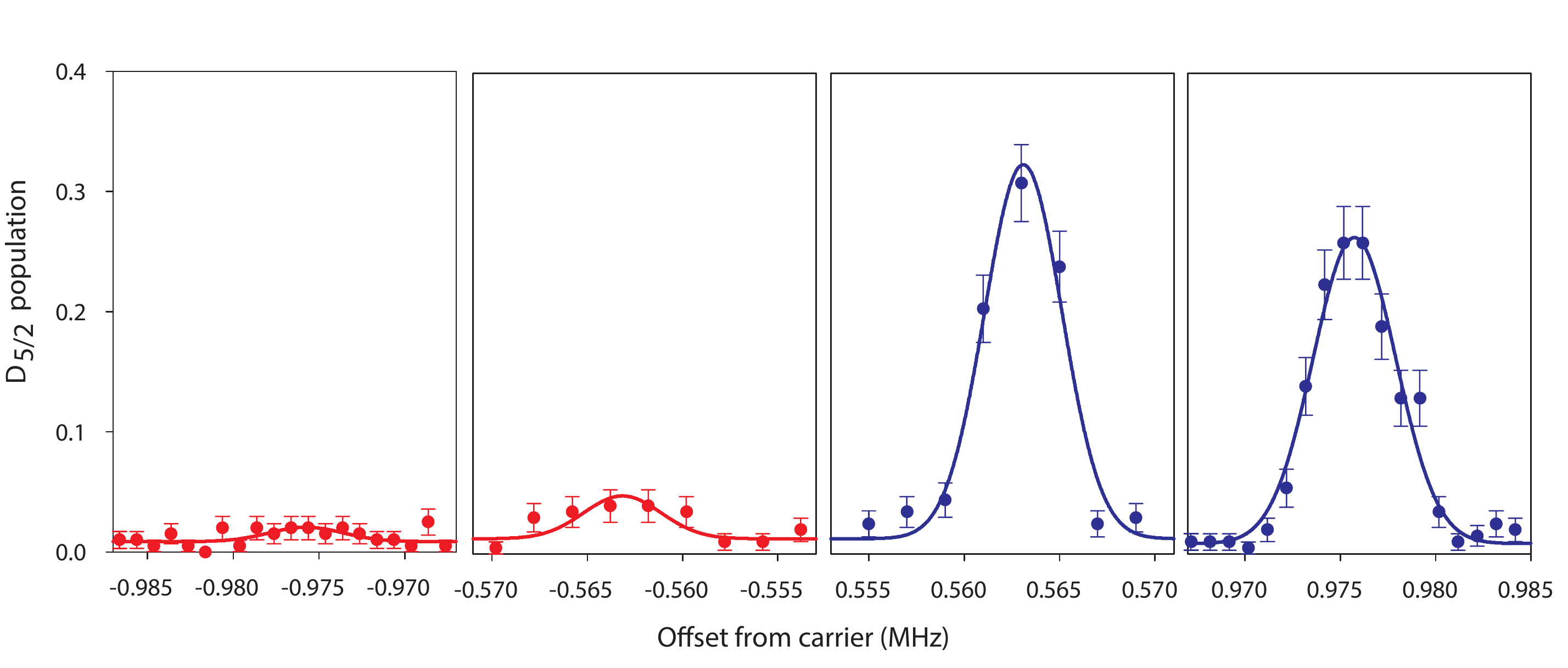}
\caption{ Measurement of the S$_{1/2}$ (m$_{j}$ = -1/2)$\longrightarrow$D$_{5/2}$ (m$_{j}$ = -5/2) first-order axial sidebands for the  $^{40}$Ca$^{+}$-  $^{40}$CaH$^{+}$ crystal by electron shelving. Red and blue sidebands are fit to a Gaussian of the same width but variable amplitude. Comparison of sideband heights yields an average mode occupation of $\overline{n}_{\mathrm{COM}}=0.13 \pm 0.03$  for the center of mass mode and $\overline{n}_{\mathrm{BM}}=0.05 \pm 0.02$ for the breathing mode. }
\label{fig:sidebands}
\end{figure*}
Before cooling the molecular ion, we optimized the ground state cooling of a single atomic ion.  After Doppler cooling, the sidebands are approximately the same height.  The temperature is measured to be approximately 0.75 mK using the carrier Rabi oscillation decoherence method~\cite{roos,rick}, which is close to the Doppler cooling limit of $T=0.53$ mK. Sideband cooling is performed on the axial mode and $\overline{n}_{\mathrm{axial}}$=0.1 is typically achieved as measured by the peak height comparison method \cite{diedrich}. Ground state cooling of the motion in three dimensions was not achieved due to an unusually high radial heating rate.  With the 729 nm laser oriented  45$^{\circ}$ off the trap axis, we were only able to cool to $\overline{n}_{\mathrm{radial}}=1$ corresponding to 50\% population in the ground state. This could be due to residual micromotion or stronger fluctuating electric fields along the radial direction than the axial direction.

Our molecular ion results are presented in Fig. \ref{fig:sidebands}. After sideband cooling, we observe that the red peak height is greatly suppressed relative to the blue peak height.  Peak height comparison reveals $\overline{n}_{\mathrm{COM}}=0.13 \pm 0.03$ and $\overline{n}_{\mathrm{BM}}=0.05 \pm 0.02$.  To determine the temperature, we match the expected occupation of the oscillator as a function of temperature with the measured occupation. We find $T_{\mathrm{COM}}=12.47 \pm 0.03~\mu$K and $T_{\mathrm{BM}}=15.36 \pm 0.01~\mu$K, which is more than a factor of 30 below the Doppler cooling limit. The presented data is typical and similar results were seen even when imperfect compensation resulted in a shifted breathing mode frequency \cite{BarrettPRA2003,goeders}.

The sideband cooling results are an important step towards the implementation of quantum logic spectroscopy \cite{wineland} or photon-recoil spectroscopy \cite{WanNatComm2013}. These techniques are limited by trap heating which acts as unwanted background that could mask the signal.  For our current experiment, we measure a single ion background heating rate of 0.1 quanta/ms comparable with other experiments in similar scale traps \cite{RoosPRL1999}. The heating rate of the atomic and molecular ion on the center of mass mode was 0.3 quanta/ms. If this mode is used for spectroscopy, the heating rate provides an idea of how quickly the molecular ion must absorb or scatter photons to have a detectable signal.

\section{Conclusions}
We have shown sympathetic sideband cooling of $^{40}$CaH$^{+}$ co-trapped with a $^{40}$Ca$^{+}$ ion. We achieve $<$16 $\mu$K translational temperature for a molecular ion by demonstrating the sideband cooling of both axial modes. The similar masses of  $^{40}$CaH$^{+}$ and $^{40}$Ca$^{+}$ maximizes the sympathetic cooling efficiency \cite{bowe}. However, this method is general and can be extended to any combination of two ions with  $\mu$ between 0.2 and 5 \cite{WubbenaPRA2012}. As an example of the range of the technique, the sympathetic ground state cooling of an amino acid (57-186 amu) by Ca$^+$ or small peptides containing up to ten amino acids by Yb$^+$ is possible.

 The sideband cooling of molecules enables the high precision spectroscopy of molecular ions required for tests of fundamental physics. Our experiments make possible the measurement of rovibrational lines of  $^{40}$CaH$^{+}$ to search for time variation in the ratio of the mass of the electron to the mass of the proton \cite{KajitaJPhysB2011}.  For these experiments, only a single, well-cooled normal mode is required, and our observation of the  near ground-state cooling of both axial modes of motion is more than sufficient.

\section*{Acknowledgments}
This work was supported by the Army Research Office (ARO)  (W911NF-12-1-0230), the National Science Foundation, Center for Chemical Innovation - Quantum Information for Quantum Chemistry (CHE-1037992),  and the Office of the Director of National Intelligence - Intelligence Advanced Research Projects Activity through ARO (W911NF-10-1-0231).

\bibliographystyle{IOPN}
\bibliography{sidebandcool-paper}

\end{document}